# Eugene Garfield's Scholarly Impact: A Scientometric Review

Chaomei Chen
College of Computing and Informatics
Drexel University

**Abstract**
The concept of citation indexing has become deeply involved in many parts of research itself and the broad environment in which research plays an integral role, ranging from research evaluation, numerous indicators, to an increasingly wider range of scientific disciplines. In this article, we pay tribute to Eugene Garfield and present a scientometric review of the intellectual assets that he brought to us. In addition, we explore the intellectual landscape that has subsequently evolved in connection to many of his ideas. We illustrate what systematic reviews of the scientific literature may reveal and what we may learn from the rich information conveyed through citation-induced patterns. The study is conducted with CiteSpace, one of many science mapping tools based on data from the Web of Science and Scopus. Without Garfield's inventions, none of these would be possible.

**Introduction**
The idea of citation indexing was originally proposed by Eugene Garfield to provide an alternative to information retrieval. When we search for relevant documents, the most straightforward strategy is to find documents that share enough similarities to our need by matching vocabularies used between these documents and a description of our need, typically, through a search query. A well-known problem is called the vocabulary mismatch in that documents can be highly relevant even if they do not share a common vocabulary. For example, if we are interested in research on scientific uncertainty, studies on how scientific consensus is reached are likely to be relevant. However, it may be hard for us to think of alternative query terms such as consensus when we concentrate on uncertainties, which would be more likely to be related to a lack of consensus. Incrementally expanding a query with semantically similar words is relatively easy. In contrast, it is much harder to think of words that may not appear to be semantically related to our original search query. In situations such as this, what could be useful would be an intellectual bridge that connects two vocabularies or more generically two entities without apparent connections. Citation indexing is one mechanism that can serve the role of such an intellectual bridge.

In a scientific publication, authors may refer to previously published articles in their text explicitly. Such explicit references are the basis of citation indexing. The primary assumption of citation indexing is that citations reflect some underlying relevance between the citing article and the cited article. The citing article is also known as the source article, whereas the cited article is known as the target article, or more commonly, the cited reference.

The basic assumption of citation analysis, or any citation-based studies, is that citations, their structural and dynamic patterns and trends, reveal something useful. Many scholars have evidently adopted the assumption or at least the spirit of it and conduct various analyses based on information and insights drawn from citations in scholarly publications. On the other hand, citation analysis has been criticized. The one that appears to have the most convincing argument is that one may never know what motivates an author makes one particular citation. Some may argue authors themselves do not necessarily know what they are doing. For example, authors may cite references they have never read, which would become evident if authors repeat even some of the same peculiar errors in how they cite references. Other criticisms include that authors tend to cite review papers more than original research articles. If we use citations to guide our search for relevant documents, the majority of these criticisms do not seem to create too much of disagreements. However, many of these criticism can be easily activated when citations are used directly or indirectly as a source of quantitative measures of intellectual merits, scholarly impacts, or other types of indicators. Does an indicator measure what it is intended to measure? How should we handle situations in which different indicators lead to conflicting or contradicting results? Many of these questions are reaching an increasingly wide range of scientific disciplines and professions. Some of them have been addressed by researchers in relevant fields for many years. Some are recently emerged and remain to be open-ended questions. Many areas of research are connected to these questions at various levels of abstraction. What they do have in common is that they are all connected in one way or another to the concept of citation.

In this article, we aim to present a systematic scientometric review of the relevant scientific literature based on two sets of publications: A) a set of 1,558 publications authored or co-authored by Eugene Garfield and B) a set of 5,054 publications that cite the set A. Set A represents Garfield's original publications, whereas Set B represents the impact of Set A through citation indexing.

**Related Work**

Garfield's scholarly impact has been a subject of study for a long time. Bensman (2007), for example, presents a historical review of Garfield and the impact factor. Leydesdorff (2010) visualized animated Garfield's oeuvre using title words, co-authors, and journal names.

Recent tributes to Garfield are made by Small (2017) and by van Raan and Wouters (2017), a Reference Publication Year Spectroscopy (RPYS) of Garfield's publications by Bornmann, Haunschild, and Leydesdorff (2017), a keyword co-occurrence analysis of the context of citations referencing Garfield's publications by Bornmann, Haunschild, and Hug 2017.

In this article, we present a scientometric review of Garfield's own publications and his scholarly impact in terms of publications that cited Garfield's publications. We use CiteSpace to generate a variety of visualizations, including dual-map overlays, alluvial flow visualizations, word-tree visualizations, and timeline visualizations (Chen 2004, Chen 2006, Chen et al. 2010, Chen et al. 2012, Chen 2017). We also demonstrate the visual exploration of a research theme by repeatedly applying the same visual analytic procedure at increasingly finer levels of granularity.

**Method**

In this article, we present the analysis of two sets of publications, sets $S_A$ and $S_B$, where $S_A$ consists of 1,558 publications authored or co-authored by Eugene Garfield and $S_B$ consists of 5,054 publications that cite $S_A$. $S_B$ consists of citation records from two sources, namely, the Web of Science and Scopus. Discrepancies in cited references between the two sources are resolved and standardized through a new method. Each of the publication sets is analyzed in terms of 1) a dual-map overlay visualization for distributions of source and target journals at the level of subject categories, 2) a word-tree of a hierarchical organization of author keywords, 3) a co-citation network in a cluster-view visualization, and 4) a timeline visualization of the evolution of specialties involved. In addition, an alluvial flow of major keywords is presented for $S_B$.

*Data Collection and Preprocessing*
**$S_A$: 1,558 Source Publications**
Garfield's ResearcherID is A-1009-2008.
According to Garfield's ResearcherID[1], there are 1,538 publications authored or co-authored by Eugene Garfield. The 1,558 records of publications in $S_A$ is used in a recently published study by Lutz Bornmann and his colleagues. Bornmann et al. (2017) analyzed Garfield's 1,558 publications using their Reference Publication Year Spectroscopy (RPYS) tool.

---

[1] http://www.researcherid.com/rid/A-1009-2008

**S$_B$: 5,054 Source Publications**

The second set of publications S$_B$ consists of 5,054 publications that cite at least one publication authored or co-authored by Garfield in S$_A$. Records are first retrieved from the Web of Science and Scopus and they are merged through a new process of integrating citation records from different sources.

We used Garfield's ResearcherID to retrieve records in the Web of Science and obtained 944 records. The difference between the 944 records and the 1,558 records in S$_A$ is due to the coverage of our institutional subscription of the Web of Science, starting from 1980. Searching with an institutional coverage of earlier years would find more records. Using the Related Records function in the Web of Science, we retrieved 3,596 records, which form a superset of publications that cite S$_A$. We searched Garfield's publications in Scopus by author name and found 233 records. The fewer number of Garfield's publications in Scopus is primarily due to the relatively shorter coverage of Scopus. The 233 publications are in turn cited by 2,615 publications in Scopus.

The search results from the Web of Science and Scopus are integrated in two steps. First, the source articles are merged based on composite keys. Given a record, its composite key consists of the first author's last name, the first letter of the first name, the year of publication, the volume number, and the first page. This design is based on several reasons. Unique identifiers from each data source cannot be used to match records from a difference source. DOIs are not always available. In contrast, information for a composite key is always available. If two records share the same composite key, then it is almost certain that they refer to the same publication, especially for journal articles and conference proceedings.

If the same publication has corresponding records in both the Web of Science and Scopus, the Web of Science record is selected in order to maximize the consistency for the longer period of coverage from the Web of Science than Scopus. Unlike the Web of Science, Scopus retains the title of a cited reference, which can be utilized in an enriched citation analysis. The article-merge step generated 5,054 records, including 3,366 (94%) from the Web of Science and 1,688 (65%) from Scopus. There were 915 overlapping records retrieved from both sources.

The second step of the integration of records from the Web of Science and Scopus is to recognize cited references that should be merged and standardized. Even within the same source, the same reference may have multiple variants. Take the highly cited paper that introduces the h-index as an example. The paper was authored by Hirsch and published in 2005 in Proceedings of the National Academy of Sciences of the United States of America (PNAS). Table 1 shows its variants in the Web of Science. The author name may appear as Hirsch J. E. or Hirsch J. The volume number is missing in two of the variants. In four types of variants, the page number is incorrect. In particular, the National Academy of Science may be abbreviated as NAT AC SCI or NATL ACAD SCI. The United States of America may be abbreviated as USA or US. Similar problems appear in Scopus (See Table 2). These variants need to be resolved and consolidated to a standardized form. The standardized form should reduce or eliminate the number of fields that have missing information.

Table 1. The same reference may have multiple variants in the Web of Science.

| Hirsch JE | 2005 | P NATL ACAD SCI USA | V102 | P16569 | DOI 10.1073/pnas.0507655102 |
|---|---|---|---|---|---|
| Hirsch JE | 2005 | P NATL ACAD SCI USA | V102 | P16569 | DOI [10.1073/pnas.0507655102, DOI 10.1073/PNAS.0507655102] |
| Hirsch J. | 2005 | P NATL ACAD SCI USA | V102 | P165 | |
| Hirsch J. E. | 2005 | P NATL ACAD SCI USA | V102 | P4 | |
| Hirsch J. E. | 2005 | P NAT AC SCI US | | | |
| Hirsch J. E. | 2005 | P NATL ACAD SCI US | | | |

Table 2. The same reference may have multiple variants in Scopus

| Author | Title | Year | Source | Volume | Issue | Page |
|---|---|---|---|---|---|---|
| Hirsch, J.E. | An index to quantify an individual's scientific research output | 2005 | Proceedings of the National Academy of Sciences of the United States of America | 102 | 46 | pp. 16569-16572 |
| Hirsch, J.E. | An index to quantify an individual's scientific | 2005 | Proceedings of the National Academy of Sciences of the United States of | 102 | | pp. 16569-16572 |

| | research output | | | America | | | |
| Hirsch, J.E. | An Index to Quantify an Individual's Scientific Research Output | | 2005 | Proceedings of the National Academy of Sciences | 102 | 46 | pp. 16569-16572 |
| Hirsch, J.E. | | | 2005 | Proc. Natl. Acad. Sci. USA | 102 | | pp. 16569-16572 |
| Hirsch, J.E. | An index to quantify an individual's scientific research output | | 2005 | Proc. Nation. Acad. Sci. USA | 102 | 46 | pp. 16569-16572 |
| Hirsch, J. | An index to quantify an individual's scientific research output | | | Proceedings of the National Academy of Sciences USA | 102 | | pp. 16569-16572 |

We use the following heuristics to merge two records that are estimated to be similar enough (e.g., greater than 80%). Given two references that are matched based on their composite keys, that implies they must have the authors with the same last name and the first letter of the first name, the same year of publication, the same volume number, and the same starting page. The next step is to compare the similarity between their source fields and determine whether they are similar enough, for example, whether Proc. Natl. Acad. Sci. USA is the same as the Proceedings of the National Academy of Sciences. We used the Jaro-Winkler distance to measure the similarity between two strings in terms of their edit distance. The Jaro distance between two words is the minimum number of single-character changes to transform one word into the other (Jaro 1989). Given two strings $s_1$ and $s_2$, their Jaro-Winkler distance d is defined as $d_j + l*p(1-d_j)$, where $d_j$ is the Jaro distance for $s_1$ and $s_2$, l is the length of common prefix at the start of the string up to four characters, and p is a constant scaling factor, which is usually set to 0.1.

The Jaro-Winkler algorithm is used to measure the similarity between author names and between source names. The other three fields must match exactly to count as a match, namely the year of publication, the volume number, and the first page. The overall similarity is the average of the 5-field similarities. Two references are considered to be the variants of the same publication if the overall similarity is greater than 0.80. Figure 1 illustrates how two variants of Moed's 2010 article published in the Journal of Informetrics are consolidated into a unified form.

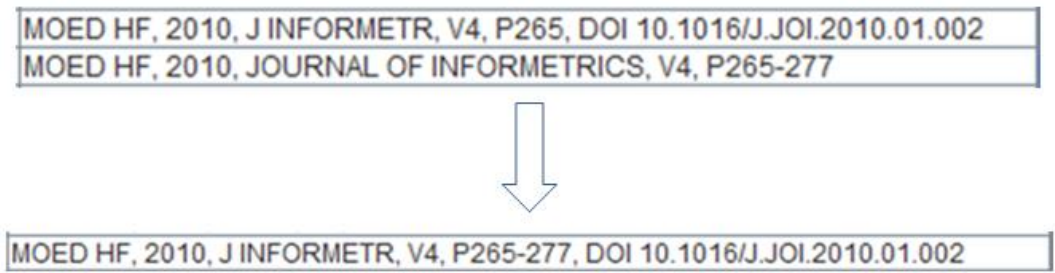

**Figure 1. Two variants of Moed's 2010 article are consolidated into a unified form.**

The 5,054 publications in $S_B$ cited 163,108 references. The consolidation procedure reduces the references to 147,585 unique references. An overview of the method is shown in Figure 2.

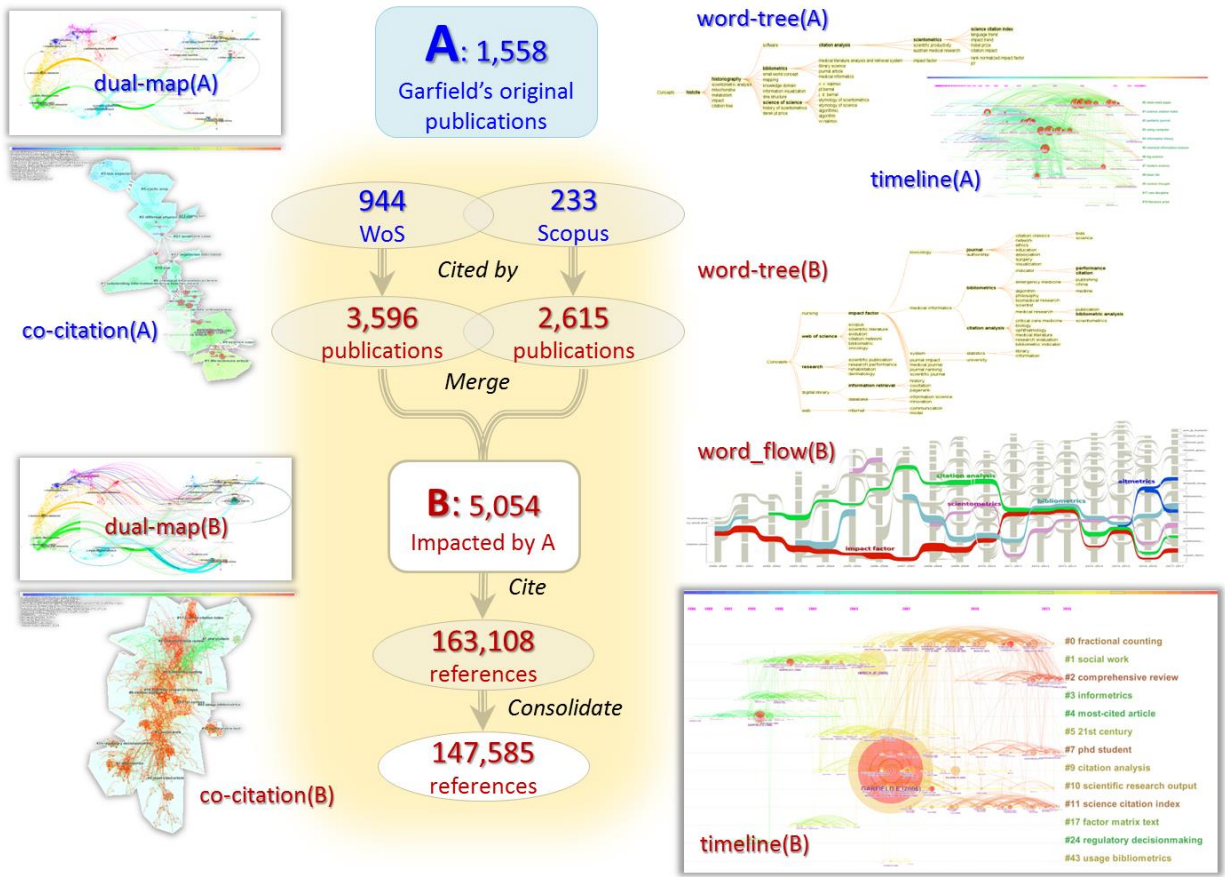

**Figure 2. An overview of the analysis.**

## Garfield's Publications (S$_A$)

### *References Cited Most By Garfield*

The publication that has been cited the most by Garfield himself is his 1972 article in Science on ranking journals based on citations. The second most cited one is his 1955 article in Science on citation indexing, the one about linking ideas through citation indexing. The third one is his 2006 publication in JAMA on the history and meaning of the journal impact factor. The fourth one is about journal impact factor. The fifth one is again on journal impact factor. See Table 3 for the top 10 references cited most by Garfield in S$_A$.

Table 3. References cited most by Garfield.

| # | Citations | Publications authored or co-authored by Garfield |
|---|---|---|
| 1. | 1283 | Garfield, E. (1972) CITATION ANALYSIS AS A TOOL IN JOURNAL EVALUATION - JOURNALS CAN BE RANKED BY FREQUENCY AND IMPACT OF CITATIONS FOR SCIENCE POLICY STUDIES. SCIENCE, 178(4060), pp. 471-+. |
| 2. | 885 | Garfield, E. (1955) CITATION INDEXES FOR SCIENCE - NEW DIMENSION IN DOCUMENTATION THROUGH ASSOCIATION OF IDEAS. SCIENCE, 122(3159), pp. 108-111. |
| 3. | 774 | Garfield, E. (2006) The history and meaning of the journal impact factor. JAMA-JOURNAL OF THE AMERICAN MEDICAL ASSOCIATION, 295(1), pp. 90-93. |
| 4. | 332 | Garfield, E. (1999) Journal impact factor: a brief review. CANADIAN MEDICAL ASSOCIATION JOURNAL, 161(8), pp. 979-980. |

| 5. | 283 | Garfield, E. (1996)  How can impact factors be improved?. BRITISH MEDICAL JOURNAL, 313(7054), pp. 411-413. |
| 6. | 277 | Garfield, E. (1979)  IS CITATION ANALYSIS A LEGITIMATE EVALUATION TOOL. SCIENTOMETRICS, 1(4), pp. 359-375. |
| 7. | 216 | Garfield, E. (1970)  CITATION INDEXING FOR STUDYING SCIENCE. NATURE, 227(5259), pp. 669-&. |
| 8. | 188 | Garfield, E. (1963) NEW FACTORS IN EVALUATION OF SCIENTIFIC LITERATURE THROUGH CITATION INDEXING. AMERICAN DOCUMENTATION, 14(3), pp. 195-&. |
| 9. | 180 | Garfield, E. (1964)  SCIENCE CITATION INDEX-NEW DIMENSION IN INDEXING - UNIQUE APPROACH UNDERLIES VERSATILE BIBLIOGRAPHIC SYSTEMS FOR COMMUNICATING + EVALUATING INFORMATION. SCIENCE, 144(361), pp. 649-&. |
| 10 | 140 | Garfield, E. (1986)  WHICH MEDICAL JOURNALS HAVE THE GREATEST IMPACT. ANNALS OF INTERNAL MEDICINE, 105(2), pp. 313-320. |

Table 4 summarizes the distributions of Garfield's publications in terms of the relatively recent Web of Science categories and the more traditional Subject Categories. Multidisciplinary Sciences is the largest group, followed by Social Sciences. The third group is much smaller, containing 227 articles in Information Science & Library Science. The fourth position is Computer Science, although there are discrepancies between the two categorization schemes.

**Table 4. Subject areas of Garfield's publications.**

| Web of Science Categories | | Subject Categories | |
|---|---|---|---|
| 1341 | Multidisciplinary Sciences | 1341 | Science & Technology - Other Topics |
| 1155 | Social Sciences, Interdisciplinary | 1157 | Social Sciences - Other Topics |
| 227 | Information Science & Library Science | 227 | Information Science & Library Science |
| 53 | Computer Science, Information Systems | 79 | Computer Science |
| 30 | Medicine, General & Internal | 30 | General & Internal Medicine |
| 29 | Computer Science, Interdisciplinary Applications | 26 | Chemistry |
| 21 | Chemistry, Multidisciplinary | 20 | Engineering |

*Dual-Map Overlays: Garfield's Publications vs His Followers' Publications*
Macroscopic views at the disciplinary level are visualized in terms of dual-map overlays. Dual-map overlays are introduced by Chen and Leydesdorff (2014). Dual-map overlays consist of a dual-map base and multiple layers of overlay visualization. The dual-map base consists of two maps of journals. On the left side is a map of source journals where an article is published. On the right side of the dual-map is a map of target journals to which references cited by the article were originally published. Given a set of publications, an overlay depicts all the resolvable references in the dataset and visualizes them as citation links from the source journal map to the target journal map. A resolvable reference means that it is published in a journal featured in the dual-map. If a cited reference is a book, then it cannot be resolved in the dual-map base and therefore it is not shown in such visualizations. One should bear in mind such omissions when interpreting these maps. To reduce the clutter caused by the excessive number of citation links across a dual-map overlay, citation links can be bundled together by aggregating links that are within the same source or target areas, e.g., within a radius of 500 pixels on the source journal map and the target journal map. Dual-map overlays highlight the predominating interdisciplinary citation links.

Figure 3 shows a dual-map overlay of Garfield's publications in $S_A$, i.e. the 1,558 publications authored or co-authored by Garfield. Citation links are bundled using the z-score function in CiteSpace. Aggregated citation paths originated in the source journal map on the left and point to target journals in the target journal map on the right. The three major clusters of source journals are journals in molecular biology and immunology (yellow), medicine and clinical journals (green), psychology, education, and health journals (cyan). Each source journal group is connected to its own counterpart in the target journal map, except that the psychology, education, health group cross-references molecular biology and genetics journals as well as journals on systems and computing. Macroscopic views at this level show that Garfield published in journals of three disciplines, echoing the most frequent subject categories his publications belong to. In addition, the dual-map overlay also reveals which disciplinary areas he cited the most.

The largest circle in the source journal map indicates that Garfield's many publications are in molecular biology and immunology journals. The largest circle in the target map is in the same disciplinary area.

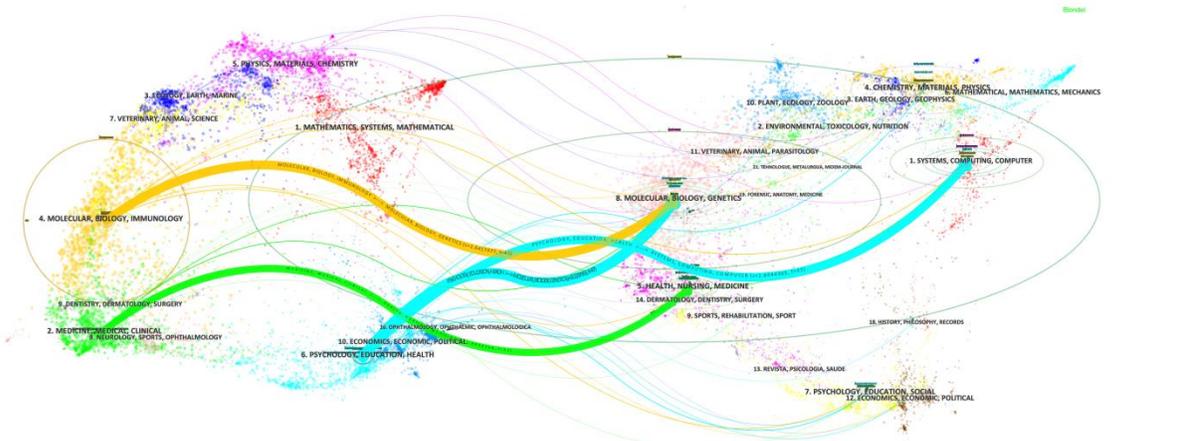

Figure 3. A dual-map overlay of Garfield's publications in $S_A$.

Figure 4 shows a dual-map overlay of Garfield's followers' 3,596 publications in the Web of Science. We did not use the 5,054 publications merged between the Web of Science and Scopus because 1) the overlapping records are already included in the Web of Science set and 2) Scopus-specific sources are not resolvable with the current base map. Generating a new base map that can accommodate both the Web of Science and Scopus would be useful but it is beyond the scope of the present study.

The four major trajectory bundles in the dual-map overlay of Garfield's publication also appear in his followers' dual-map overlay visualization. In addition, a new trajectory from the medicine and clinical source journals links to the molecular biology and genetics target journal group. Another difference is a linkage from the economics and politics source journals to target journals in the same discipline.

The largest circle on the left is centered at the source journal cluster labeled as psychology, education, health. This area contains information science and library science journals. The largest circle on the right is located at the areas of journals on computing and systems. As it appears, a major development has taken place in connections between information science and computing systems.

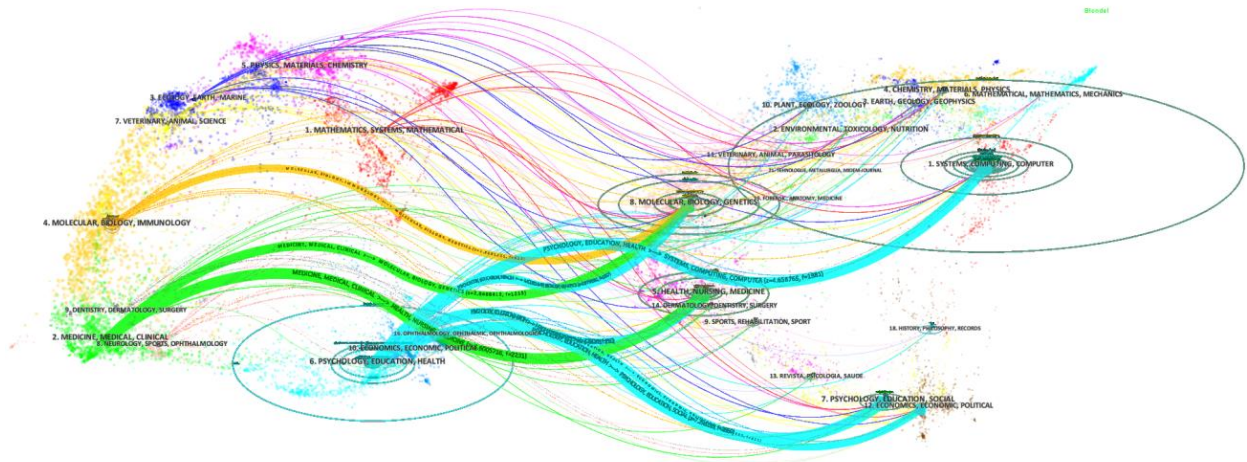

**Figure 4. A dual-map overlay of Garfield's followers' 3,596 publications in the Web of Science. Citation links are bundled by z-scores.**

Figure 5 shows the same dual-map overlay without bundling citation links. Links are colored by their source journal clusters' colors.

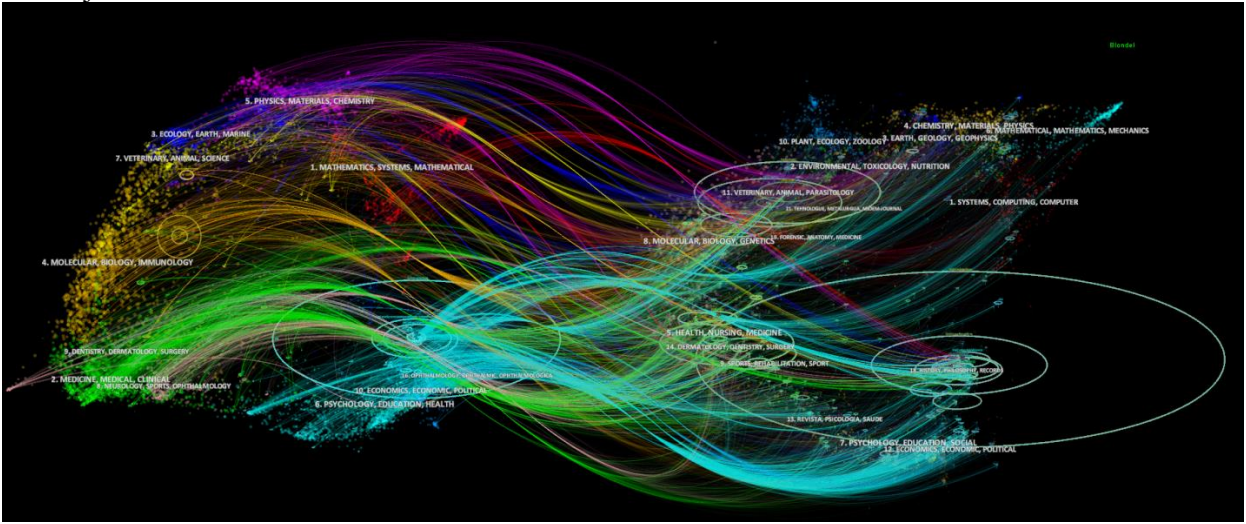

**Figure 5. A dual-map overlay of Garfield's followers' 3,596 publications without bundling citation links.**

*Word-Trees of Author Keywords*

Given a set of publications containing keywords provided by their original authors, a word-tree visualization represents a hierarchy of keywords derived from their co-occurrence patterns. A hierarchy is generated using the method introduced by Tibély et al. (2013). Figure 6 shows the distribution of distinct keywords (DE) in $S_A$. Figure 7 shows a word-tree visualization derived from these author keywords.

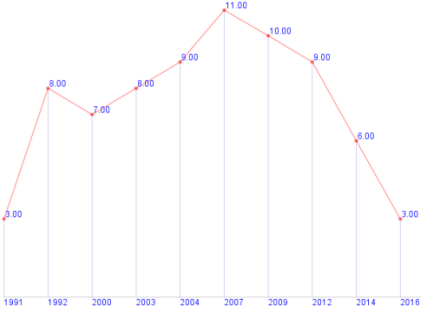

**Figure 6. The number of author keywords (DE) per year in $S_A$.**

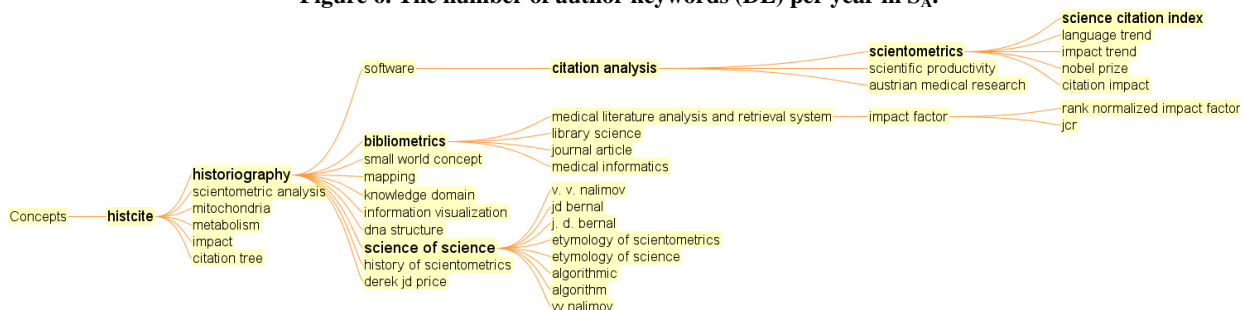

**Figure 7. A word-tree of author keywords in Garfield's 1,558 publications.**

HistCite is the leading keyword. The top branch contains the most salient path, which characterizes HistCite with keywords such as historiography, software, citation analysis, scientometrics, and science citation index. Two smaller branches are under the historiography: bibliometrics and science of science.
Figure 8 is a hybrid word-tree containing both Web of Science categories with three or more counts and all the author keywords. The Multidisciplinary Sciences branches to two categories of information science and social sciences.

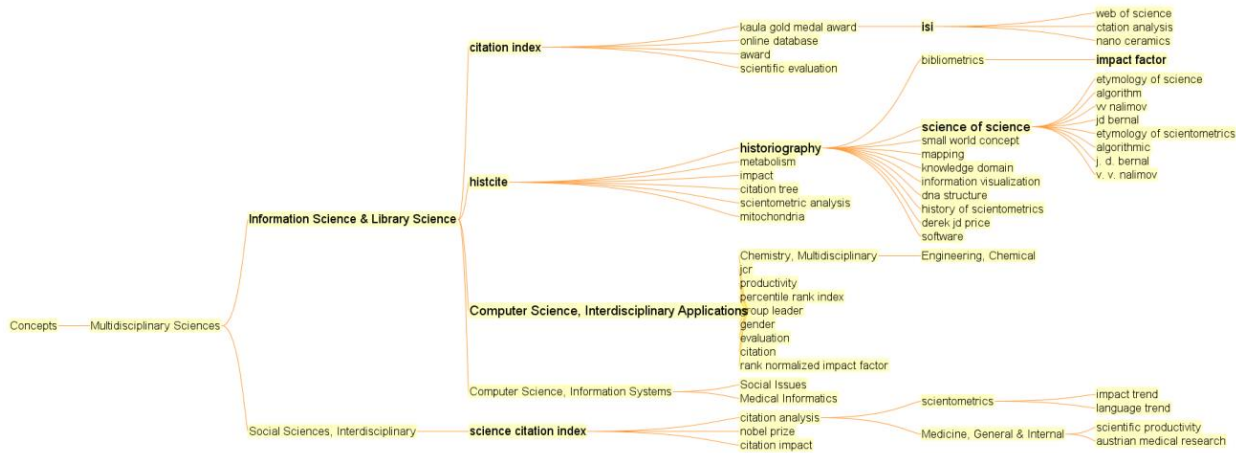

**Figure 8. A word-tree of the Web of Science categories (WC≥3) and author keywords (DE≥1).**

## Co-Citation Clusters ($S_A$)

The current method of co-citation analysis was originated from Henry Small. If an article cites references $r_a$ and $r_b$, then the two references are co-cited. A co-citation relation is local information without involving other nodes. However, it provides an effective mechanism to aggregate local relations and form a network that can represent a global structure.

Table 5 lists the major co-citation clusters based on citations made by the 1,558 publications in $S_A$. Each area is a cluster within which member references are frequently co-cited. Each cluster is labeled by title words in articles that cite them using log-likelihood ratio tests (LLR). For example, cluster labels such as #0 most-cited paper and #1 science citation index are phrases from the titles of Garfield's publications that cite these references.

**Table 5. Co-citation clusters based on citations made by publications in $S_A$.**

| Cluster | Size | Silhouette | Mean(Year) | Log-Likelihood Ratio (LLR) |
|---|---|---|---|---|
| 0 | 62 | 0.642 | 1980 | Most-cited paper |
| 1 | 66 | 0.716 | 1967 | Science citation index |
| 2 | 49 | 0.781 | 1970 | Pediatric journal |

| 3 | 41 | 0.739 | 1975 | Using computer |
| 4 | 35 | 0.910 | 1961 | Information theory |
| 5 | 34 | 0.861 | 1972 | Chemical information science |
| 6 | 36 | 0.902 | 1974 | Big science |
| 7 | 28 | 0.876 | 1981 | Modern science |
| 8 | 18 | 0.917 | 1976 | Basic list |
| 9 | 18 | 0.842 | 1989 | Random thought |
| 17 | 5 | 0.999 | 1959 | New discipline |
| 19 | 4 | 0.994 | 1987 | Literature prize |

**Cluster View Visualization**

Figure 9 shows a cluster view of a co-citation network derived from the 1,558 publications in $S_A$. The colors of these clusters represent the average year of publication. Red tree rings represent periods of citation burst.

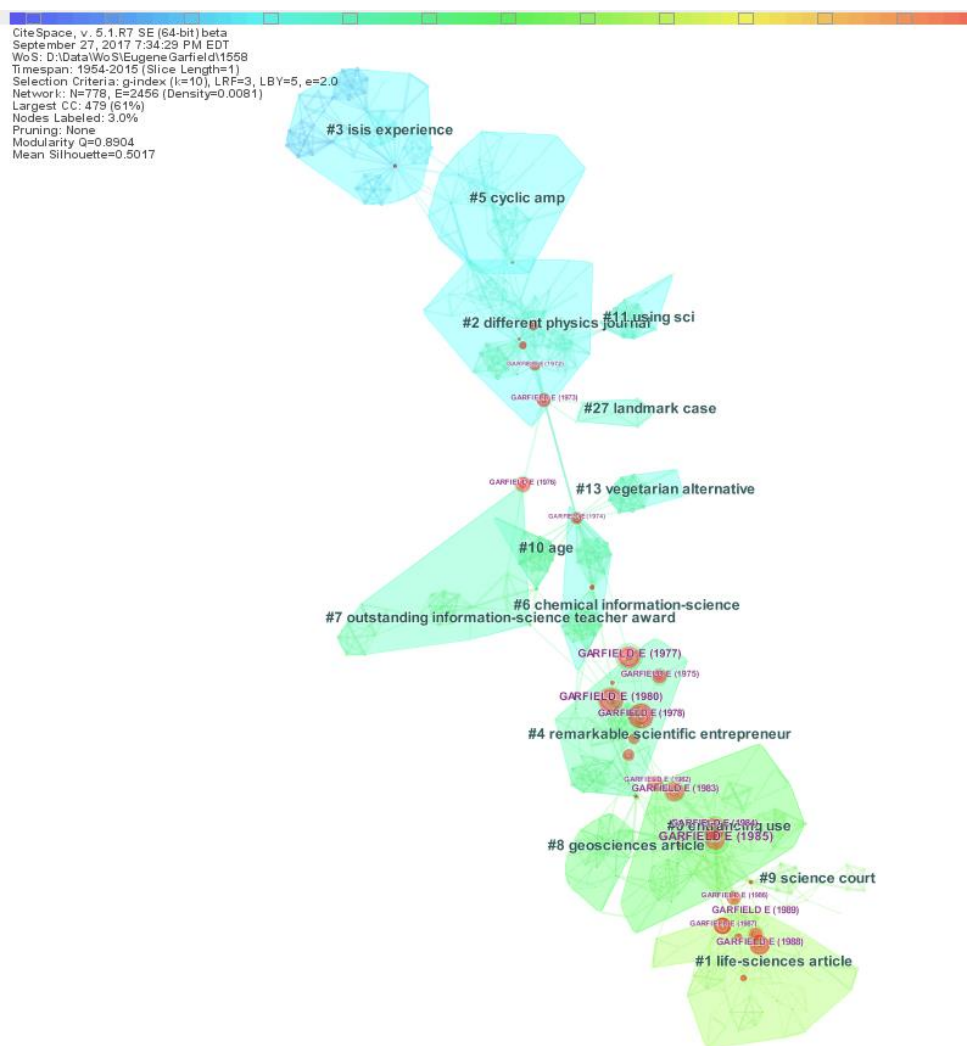

**Figure 9. A co-citation network generated from 1,558 publications in $S_A$. Red tree rings show periods of citation burst.**

**Timeline Visualization**

Figure 10 shows a timeline visualization of the co-citation clusters. Each timeline runs from the left to the right. Clusters are shown from the top downwards in the descending order of their size. The earliest co-citation cluster is Cluster #1 science citation index, starting from 1955. Around 1970, the major attention was shifted to Cluster #2, which in turn shifted to Cluster #3 after 1975. Cluster #3 contains a series of highly cited articles and sustained periods of citation burst. The formation of Cluster #0 was long before its sustained stream of highly cited articles between 1980s and 1990s.

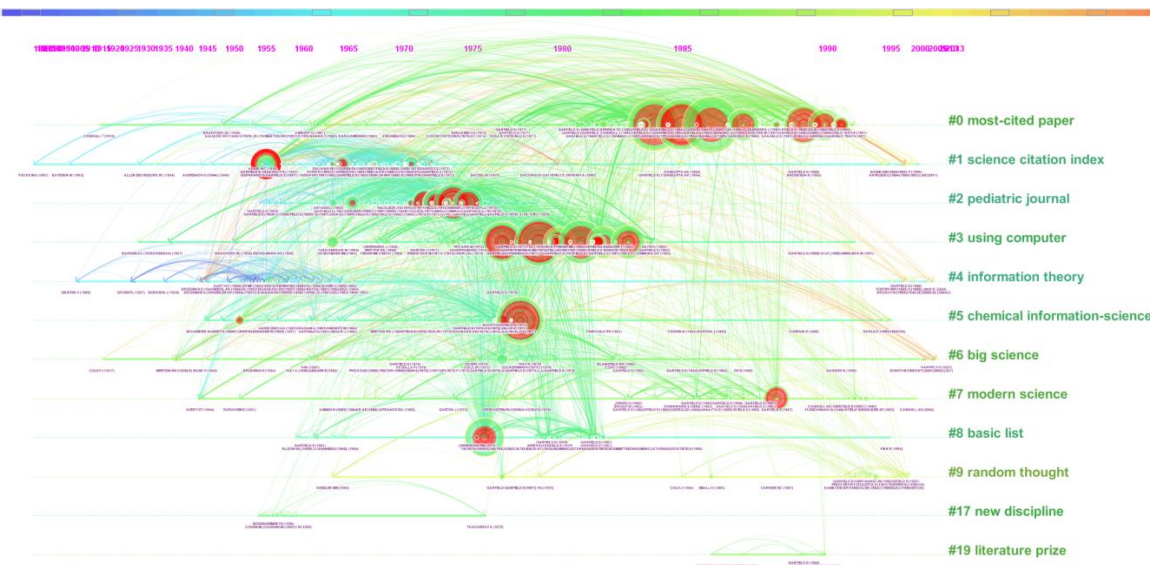

**Figure 10. A timeline visualization of co-citation clusters through a fisheye lens with citation burst for four years or more.**

Table 6 shows the most cited references in Cluster #1. Garfield's 1955 publication in Science is a groundbreaking paper with a citation half-life of 33 years.

**Table 6. Highly cited references in Cluster #1.**

| Citations | Burst | Author | Year | Source | Volume | Page | Half-Life |
|---|---|---|---|---|---|---|---|
| 48 | 6.12 | Garfield E | 1955 | SCIENCE | 122 | 108 | 33 |
| 11 | 4.90 | Garfield E | 1963 | AM DOC | 14 | 195 | 9 |
| 22 | 9.83 | Garfield E | 1964 | SCIENCE | 144 | 649 | 5 |
| 19 | 5.56 | Garfield E | 1970 | NATURE | 227 | 669 | 3 |

Table 7 shows that the 1965 Science article by DJD Price was the highest cited article except those of Garfield's own ones. The Science article by Price has a citation half-life of 11 years. Garfield's own 1972 Science article has a citation half-life of four years.

**Table 7. Highly cited references in Cluster #2 ranked by citation half-life.**

| Citations | Burst | Author | Year | Source | Volume | Page | Half-Life |
|---|---|---|---|---|---|---|---|
| 16 | 4.31 | Price DJD | 1965 | SCIENCE | 149 | 510 | 11 |
| 36 | 11.69 | Garfield E | 1972 | SCIENCE | 178 | 471 | 4 |
| 33 | 11.92 | Garfield E | 1974 | CURR CONTENTS | 0 | 5 | 3 |
| 8 | 4.85 | Garfield E | 1970 | CURRENT CONTENT 0304 | 0 | 4 | 3 |
| 42 | 15.82 | Garfield E | 1973 | CURR CONTENTS | 0 | 5 | 2 |
| 30 | 12.70 | Garfield E | 1972 | CURRENT CONTENT 1101 | 0 | 5 | 2 |
| 29 | 10.90 | Garfield E | 1971 | CURR CONTENTS | 0 | 5 | 2 |
| 9 | 5.31 | Garfield E | 1973 | CURRENT CONTENT 0207 | 0 | 7 | 2 |

**The Impact of Garfield's Work ($S_B$)**

The impact of Garfield's work is reflected by the breadth and the depth of 5,054 publications in $S_B$. The predominant document type of the 5,054 publications is the Article type (Table 8). The second largest type is Editorial Material, followed by other types such as Review and Letter.

Table 8. Document types in $S_B$.

| Count | Document Type |
|---|---|
| 4109 | Article |
| 431 | Editorial Material |
| 224 | Review |
| 136 | Letter |
| 105 | Article; Proceedings Paper |
| 10 | Note |
| 10 | Book Review |
| 9 | Correction |
| 8 | Reprint |
| 3 | Item About an Individual |
| 2 | Biographical-Item |
| 1 | Discussion |
| 1 | News Item |
| 1 | Correction, Addition |
| 1 | Review; Book Chapter |

*Most Cited Publications in $S_B$*

The most cited publications in $S_B$ are identified based on the TC field in the Web of Science or the counterpart field in Scopus, whichever is higher (Table 9). The top-10 list includes two of Garfield's own publications, the 2006 JAMA paper on the history and meaning of the journal impact factor and the 1999 paper, which is also on the topic of journal impact factor. A few publications on the list are beyond the information science discipline. Publications that belong to information science include Bornmann's 2008 review on citing behavior and Meho's 2007 comparisons of the Web of Science, Scopus, and Google Scholar. In particular, the list contains three science mapping publications, namely, Chen's 2006 article on CiteSpace, van Eck's 2010 article on VOSviewer, and Boyack's 2005 article on mapping the backbone of science.

Table 9. Most Cited Publications in $S_B$. References may have multiple authors, but only the first authors shown.

| Citations | Highly Cited Publications in $S_B$ |
|---|---|
| 1528 | Kluger, AN. (1996) The effects of feedback interventions on performance: A historical review, a meta-analysis, and a preliminary feedback intervention theory. PSYCHOLOGICAL BULLETIN, 119(2), pp. 254-284. |
| 846 | Garfield, E. (2006) The history and meaning of the journal impact factor. JAMA-JOURNAL OF THE AMERICAN MEDICAL ASSOCIATION, 295(1), pp. 90-93. |
| 758 | Munns, R. (1993) PHYSIOLOGICAL PROCESSES LIMITING PLANT-GROWTH IN SALINE SOILS - SOME DOGMAS AND HYPOTHESES. PLANT CELL AND ENVIRONMENT, 16(1), pp. 15-24. |
| 420 | Chen, CM. (2006) CiteSpace II: Detecting and visualizing emerging trends and transient patterns in scientific literature. JOURNAL OF THE AMERICAN SOCIETY FOR INFORMATION SCIENCE AND TECHNOLOGY, 57(3), pp. 359-377. |
| 392 | Meho, Lokman. (2007) Impact of data sources on citation counts and rankings of LIS faculty: Web of science versus scopus and google scholar. JOURNAL OF THE AMERICAN SOCIETY FOR INFORMATION SCIENCE AND TECHNOLOGY, 58(13), pp. 2105-2125. |

| 374 | Bornmann, L. (2008) What do citation counts measure? A review of studies on citing behavior. JOURNAL OF DOCUMENTATION, 64(1), pp. 45-80. |
|---|---|
| 361 | Garfield, E. (1999) Journal impact factor: a brief review. CANADIAN MEDICAL ASSOCIATION JOURNAL, 161(8), pp. 979-980. |
| 346 | Van, Eck. (2010) Software survey: VOSviewer, a computer program for bibliometric mapping. Scientometrics, 84(2), pp. 523-538. |
| 345 | Van spall, Harriette. (2007) Eligibility criteria of randomized controlled trials published in high-impact general medical journals - A systematic sampling review. JAMA-JOURNAL OF THE AMERICAN MEDICAL ASSOCIATION, 297(11), pp. 1233-1240. |
| 334 | Boyack, KW. (2005) Mapping the backbone of science. SCIENTOMETRICS, 64(3), pp. 351-374. |

Table 10 lists top-10 references that are cited most by publications in $S_B$. Most of them are authored by Garfield, except one that introduces the h-index and Seglen's 1997 article on why the journal impact factors should not be used for research evaluation.

Table 10. Top-10 references cited most by publications in $S_B$.

| 773 | Garfield E, 2006, JAMA-J AM MED ASSOC, V295, P90, DOI 10.1001/jama.295.1.90 |
|---|---|
| 407 | Hirsch JE, 2005, P NATL ACAD SCI USA, V102, P16569, DOI 10.1073/pnas.0507655102 |
| 357 | Seglen PO, 1997, BRIT MED J, V314, P498 |
| 334 | Garfield E, 1999, CAN MED ASSOC J, V161, P979 |
| 332 | Garfield E, 1972, SCIENCE, V178, P471-479 |
| 313 | GARFIELD E, 1955, SCIENCE, V122, P108, DOI 10.1126/science.122.3159.108 |
| 293 | GARFIELD E, 1972, SCIENCE, V178, P471, DOI 10.1126/science.178.4060.471 |
| 280 | Garfield E, 1955, SCIENCE, V122, P108-111 |
| 274 | Garfield E, 1996, BRIT MED J, V313, P411 |
| 149 | GARFIELD E, 1980, CURR CONTENTS, P5 |

*Alluvial Flow of Author Keywords*

Alluvial flow is a visualization method of multiple networks over time or other sequences. We generated a network of co-occurring author keywords in each year in CiteSpace and imported these annual networks into an Alluvial Flow generator. Flows such as impact factor, citation analysis, and citation analysis almost last the entire course between 2000 and 2017 (Figure 11). Altmetrics emerged in the networks since 2015.

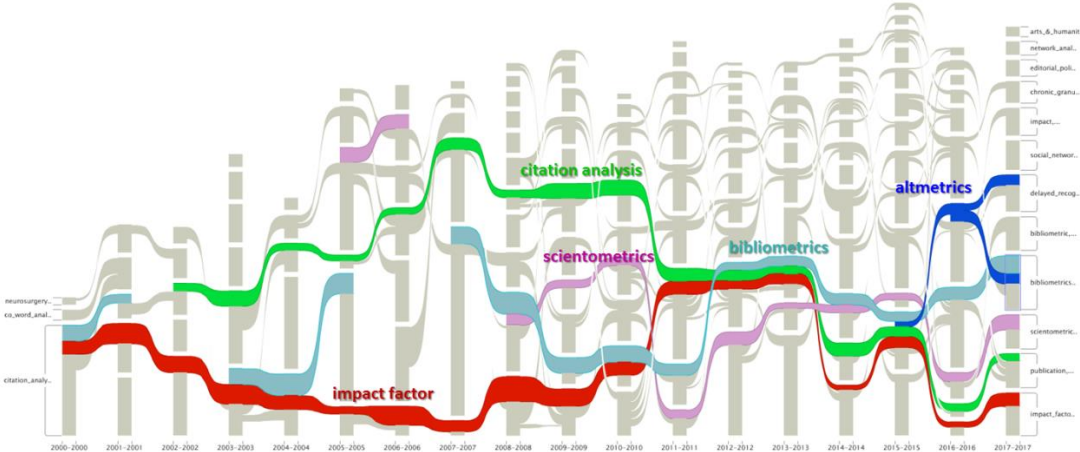

**Figure 11. An alluvial flow of author keywords (DE), selected by g-index (k=30) per year.**

## Word-Trees of Author Keywords

Figure 12 shows a few word-trees of author keywords generated with various frequency thresholds. The one at the upper left is the largest tree that contains all the author keywords available in the dataset. One can explore it interactively by zooming in and out, but it is too big to fit to a journal page. The next one consists of author keywords that appear in 20 or more publications in $S_B$. The two further to the right are generated with threshold values of 30 and 40, respectively. The one at the bottom is generated with a frequency threshold of 100, which contains keywords that appear in 100 or more publications in the dataset. The most salient path is the one that moves along the top of the tree, connecting nursing, citation, indicator, bibliometrics, emergency medicine, and publishing. Another path branches off from indicator to include performance, h index, and trend, which clearly reveals the underlying theme.

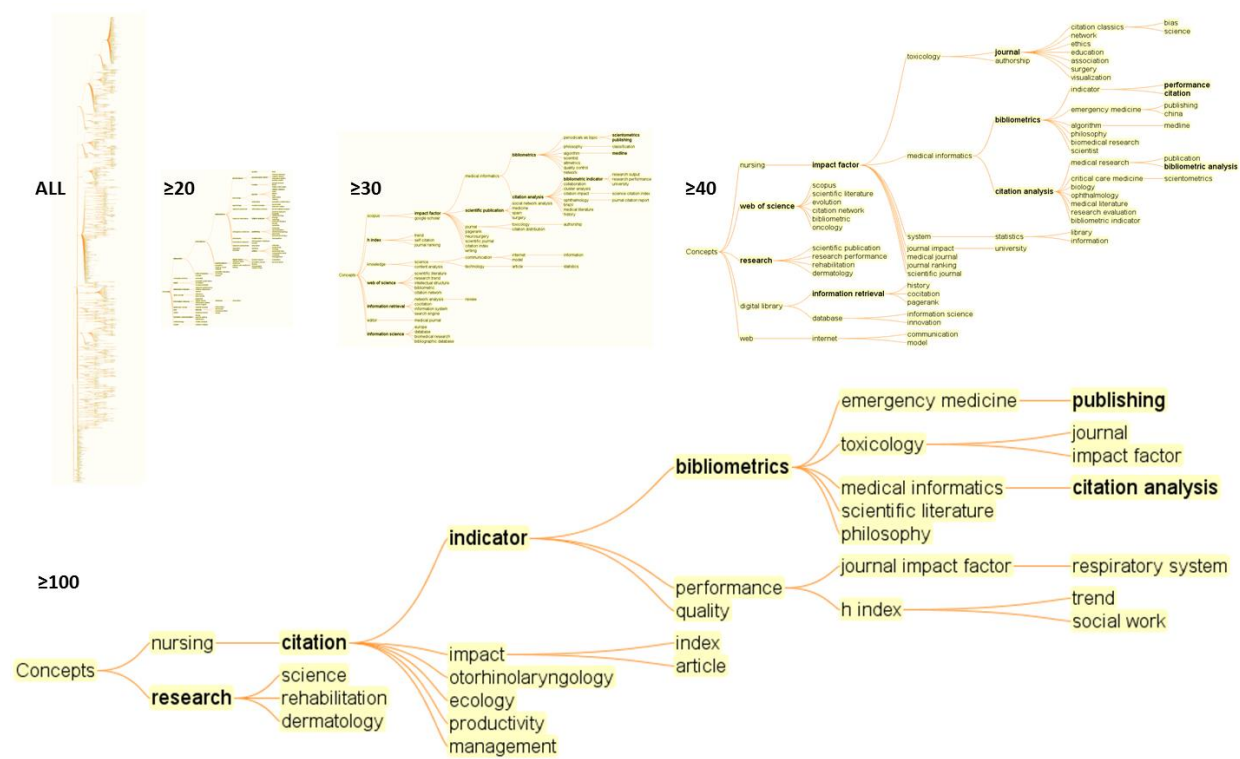

**Figure 12. Word-trees of author keywords with various frequency thresholds.**

## Co-Citation Clusters ($S_B$)

We generated a co-citation network in CiteSpace with the 5,054 publications in the impact set $S_B$. The configuration of the network was based on the g-index with a scaling factor of 30. The network consists of 2,275 cited references and 9,135 co-citation links. The largest connected component in the network contains 1,239 nodes, which is 54% of the entire network. Table 11 shows the co-citation clusters identified in the network. The largest 6 clusters all have more than 100 members each. Their silhouette scores indicate a high level of homogeneity within these clusters. In terms of the average age of a cluster, the oldest ones are Clusters #6 and #21. The most recent one is Cluster 2, with 2013 as the average year of publication. The average year of publication of Cluster #0, the largest one, is 2009.

**Table 11. Co-citation clusters from $S_B$.**

| Cluster # | Size | Silhouette | Average(Year) | LSI | LLR |
|---|---|---|---|---|---|

| 0 | 218 | 0.742 | 2009 | Impact factor | Fractional counting |
| 1 | 191 | 0.800 | 2001 | Impact factor | Social work |
| 2 | 139 | 0.909 | 2013 | Impact | Comprehensive review |
| 3 | 110 | 0.954 | 1998 | Information science | informetrics |
| 4 | 104 | 0.899 | 1994 | Impact factor | Most-cited article |
| 5 | 101 | 0.880 | 2004 | Science | 21st century |
| 7 | 99 | 0.937 | 2011 | Journals | PhD student |
| 9 | 91 | 0.883 | 2006 | Impact factor | Citation analysis |
| 10 | 82 | 0.917 | 2007 | h-index | Scientific research output |

The modularity of the network is 0.865, which means that the co-citation structure can be divided into relatively independent groups. Garfield's 2006 JAMA article is the most cited reference by the set of publications merged from the Web of Science and Scopus. The burst detection found 26 references with bursts that lasted for 6 years or longer.

Figure 13 shows a visualization of the co-citation network. The color of a link represents the year when the co-citation relation was found for the first time in the dataset. In this visualization, co-citation links made in the earlier years are located towards the bottom, whereas connections made in more recent years are located towards the top. A node with one or more tree rings in red indicates that the corresponding reference has a period of citation burst. In this case, we focus on sustained citation bursts of 6 years or more. Several Garfield's publications have such citation bursts, which is not a surprise because the entire dataset was collected based on the idea of citation indexing with reference to Garfield's publications. Garfield's publications in 1996, 1999, 2006, and 2007 are particularly highly cited by publications in $S_B$. Next to Garfield_2006 is another highly influential article by Hirsch (2005), which is the one that introduced the now widely used and, perhaps also misused, indicator of the performance and productivity of a scientist – the h-index.

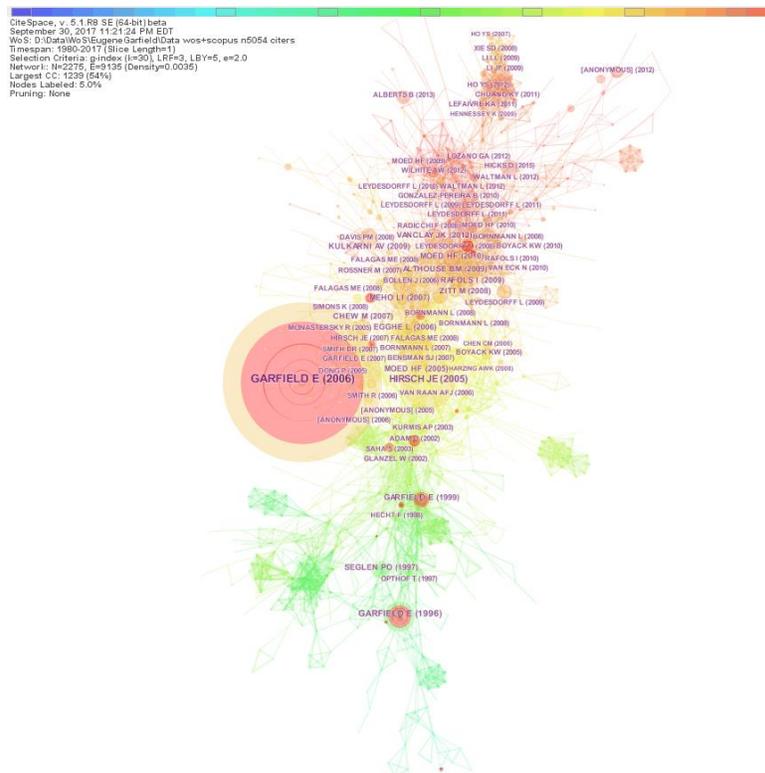

**Figure 13. A co-citation network generated by citations made by 5,054 publications in $S_B$.**

Table 12 lists references with citation bursts of 6 years or longer. The red bar represents the duration of a citation burst as well as the starting and ending years of its citation burst. Seventeen of the articles with a burst of this scale are Garfield's publications, including his publications in Current Contents in the 1980s and Scientists in late 1990s. His 1996 article in British Medical Journal had a particularly strong citation burst with the title "How can impact factors be improved?" Garfield's next article with a strong burst is a brief 2-page review of journal impact factor published in the Canadian Medical Association Journal (Garfield 1999), in which he acknowledged that the impact factor had become the subject of widespread controversy and that it might be misused in the wrong hands. Garfield reiterated his key points in 2006 in a 4-page commentary published in JAMA – the Journal of the American Medical Association. The 2006 article has the strongest citation burst and it currently has 2,044 citations on Google Scholar.

Apart from the citations on publications by Garfield, several publications by others have also attracted substantial citations and citation bursts as shown in shaded rows in Table 12. In particular, Small and Sweeney (1985) proposed two methodological improvements, namely, fractional citation counting and variable level clustering with a maximum cluster size limit. Fractional citation counting may reduce the bias toward fields such as biomedicine and biochemistry. Variable level clustering increases recalls in terms of the percentage of highly cited references included in clusters. They attributed the idea of comprehensive maps of science to Derek Price.

Moed and van Leeuwen (1996) published a correspondence in *Nature* with a provocative title: Impact factors can mislead. They started with a question on what counts as a citable item in the calculation of a journal's impact factor and ended with scenarios of how journal editors may boost their own journals' ranking regardless of any scholarly improvements.

Fassoulaki et al. (2000) published a commentary on how self-citations in six anesthesia journals affect the impact factor. In a news feature piece in *Nature*, Adam (2002) reported his interviews with bibliometric researchers concerning the misuse of citation analysis in the wrong hands and the ending of the "romantic period" for bibliometric research after Thomson bought Garfield's ISI. The "romantic period" refers to Garfield's personal interest and support in bibliometric research.

Saha et al. (2003) reported a strong correlation between the journal impact factors of nine general medical journals and a survey of clinical practitioners and researchers on the quality of these journals.

Bensman (2007) provided a comprehensive review of the political, social, and intellectual influences affecting Garfield in his creation of the impact factor based on Garfield's own writings.

Falagas et al. (2008) compared four databases as sources of biomedical publications, namely, PubMed, the Web of Science, Scopus, and Google Scholar. Their comparison aims to evaluate the usefulness of these databases for biomedical information retrieval and citation analysis.

More recent citation bursts include a 2011 article by Waltman et al. on a new indicator of research performance based on a well-known indicator developed at Leiden University, the crown indicator. The new indicator, i.e. a new crown indicator, normalizes citation counts across different fields. Leydesdorff and Bornmann (2011) addressed how fractional counting of citations affects the impact factor.

In terms of the strength of citation burst, the research community has been extensively concerned with the fairness of making comparisons across disciplines (Small and Sweeney 1985, Waltman et al. 2011, Leydesdorff and Bornmann 2011) and improvements on the impact factor to make it a useful but not dangerous tool (Moed and van Leeuwen 1996, Adam 2002).

In contrast to the controversies surrounding the use or misuse of the journal impact factor, the other fundamental invention of Garfield, citation indexing, has made profound and remarkably quiet advances. The only high-profile article on the Web of Science is the comparison made with PubMed, Scopus, and Google Scholar (Falagas et al. 2008). As it appears, it is hard to measure the impact of the idea of science citation index, but the increasing number of large-scale resources of scientific publications provide citations as integral parts of their data and services, notably Microsoft Academic, ResearchGate, Google Scholar, Scopus, and, of course, the Web of Science.

**Table 12. References with citation bursts for 6 years or longer.**

| References | Year | Strength | Begin | End | 1980–2017 |
|---|---|---|---|---|---|

| References | Year | Strength | Begin | End | |
|---|---|---|---|---|---|
| GARFIELD E, 1980, CURR CONTENTS, V0, P5 | 1980 | 28.3687 | **1980** | 1985 | |
| GARFIELD E, 1980, LIBR QUART, V50, P40 | 1980 | 6.9036 | **1980** | 1985 | |
| GARFIELD E, 1982, CURR CONTENTS, V0, P5 | 1982 | 22.6048 | **1982** | 1987 | |
| GARFIELD E, 1983, CURR CONTENTS, V0, P5 | 1983 | 35.3247 | **1983** | 1988 | |
| SMALL H, 1985, SCIENTOMETRICS, V7, P391, DOI | 1985 | 6.5476 | **1985** | 1990 | |
| GARFIELD E, 1986, ANNALS OF INTERNAL MEDICINE, V105, P313-320 | 1986 | 12.241 | **1986** | 1991 | |
| GARFIELD E, 1986, CURR CONTENTS, V0, P3 | 1986 | 22.6965 | **1986** | 1991 | |
| GARFIELD E, 1987, JAMA-J AM MED ASSOC, V257, P52, DOI | 1987 | 8.0682 | **1987** | 1992 | |
| GARFIELD E, 1987, CURR CONTENTS, V0, P3 | 1987 | 24.2875 | **1987** | 1992 | |
| GARFIELD E, 1988, CURR CONTENTS, V0, P3 | 1988 | 27.6457 | **1988** | 1993 | |
| GARFIELD E, 1990, JAMA-J AM MED ASSOC, V263, P1424, DOI | 1990 | 6.4168 | **1990** | 1995 | |
| GARFIELD E, 1996, BRITISH MEDICAL JOURNAL, V313, P411-413 | 1996 | 44.0751 | **1996** | 2001 | |
| MOED HF, 1996, NATURE, V381, P186, DOI | 1996 | 7.4982 | **1996** | 2001 | |
| GARFIELD E, 1998, SCIENTIST, V12, P11 | 1998 | 4.1849 | **1998** | 2003 | |
| GARFIELD E, 1998, SCIENTIST, V12, P12 | 1998 | 4.7834 | **1998** | 2003 | |
| GARFIELD E, 1999, CANADIAN MEDICAL ASSOCIATION JOURNAL, V161, P979-980 | 1999 | 30.847 | **1999** | 2004 | |
| FASSOULAKI A, 2000, BRITISH JOURNAL OF ANAESTHESIA, V84, P266-269 | 2000 | 10.4045 | **2000** | 2005 | |
| ADAM D, 2002, NATURE, V415, P726-729 | 2002 | 19.2553 | **2002** | 2007 | |
| GARFIELD E, 2003, JOURNAL OF THE AMERICAN SOCIETY FOR INFORMATION SCIENCE AND TECHNOLOGY, V54, P400-412 | 2003 | 6.8212 | **2003** | 2008 | |
| SAHA S, 2003, JOURNAL OF THE MEDICAL LIBRARY ASSOCIATION, V91, P42-46 | 2003 | 17.3605 | **2003** | 2008 | |
| GARFIELD E, 2006, JAMA: JOURNAL OF THE AMERICAN MEDICAL ASSOCIATION, V295, P90-93 | 2006 | 136.0356 | **2006** | 2011 | |
| GARFIELD E, 2007, INTERNATIONAL MICROBIOLOGY, V10, P65-69, DOI | 2007 | 8.199 | **2007** | 2012 | |
| BENSMAN SJ, 2007, ANNUAL REVIEW OF INFORMATION SCIENCE AND TECHNOLOGY, V41, P93-155 | 2007 | 9.4183 | **2007** | 2012 | |
| FALAGAS ME, 2008, FASEB JOURNAL, V22, P338-342, DOI | 2008 | 8.0032 | **2008** | 2013 | |
| WALTMAN L, 2011, J INFORMETR, V5, P37, DOI | 2011 | 6.4524 | **2011** | 2017 | |
| LEYDESDORFF L, 2011, JOURNAL OF THE AMERICAN SOCIETY FOR INFORMATION SCIENCE AND TECHNOLOGY, V62, P217-229 | 2011 | 10.9885 | **2011** | 2017 | |

Figure 14 shows visualizations of the network of references cited by publications in $S_B$. The one on the left highlights the temporal patterns of the growth of the network, from the earlier intellectual contributions made by Garfield in the blue and green regions, through the regions in yellow and eventually to the areas form by recent publications. The one on the right shows how much of the intellectual space is covered by the Web of Science (red) and what is added exclusively by Scopus (green). Integrating citation data from multiple sources such as the Web of Science and Scopus raises practical issues. For example, if the same reference $r$ appears in both sources, how should the two variants $r_a$ and $r_b$ be consolidated? If $r_a$ is cited by a set of publication $C_a(r_a)$ and $r_b$ is cited by a set of publication $C_b(r_b)$, then the best solution is to merge the two sets of publication $C_a(r_a) \cup C_b(r_b)$ for $r$. However, this seemingly simple operation requires an access to the entirety of both sources. The green area reminds us the coverage that may be missed out if we do not incorporate records from Scopus in this particular case. In general, it is likely that it will take both sources to achieve an adequate coverage of a topic of interest. Such issues should be addressed in further studies.

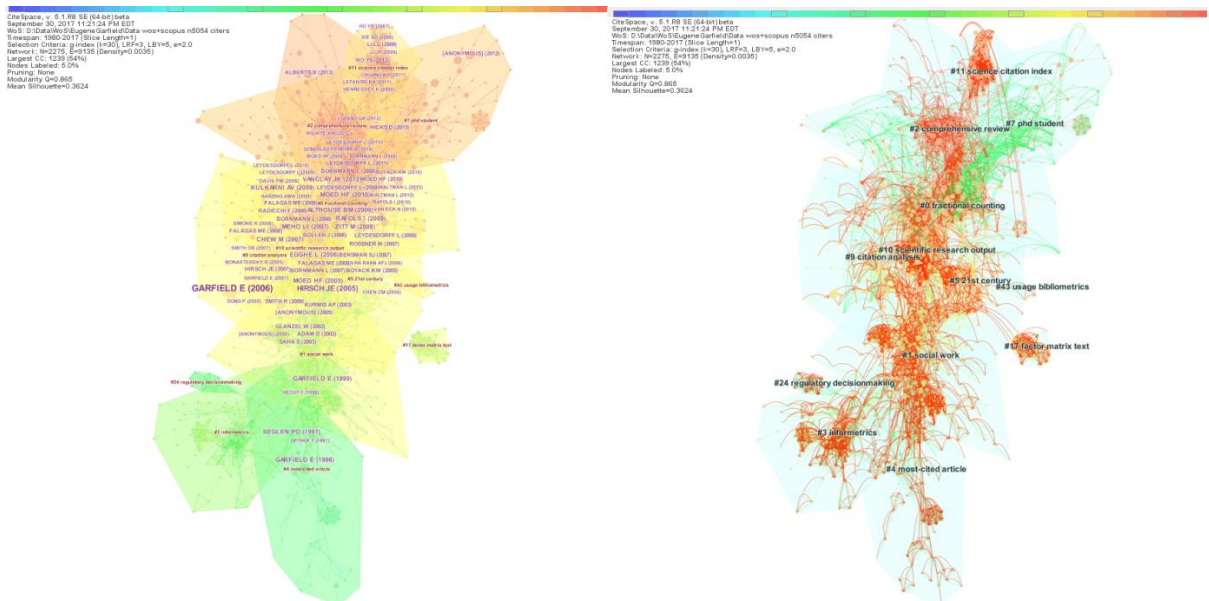

**Figure 14. Cluster views of the co-citation network. Left: Clusters colored by the average years of publication. Right: the contrast between citations from the Web of Science or overlapping records with Scopus (red) and from records exclusively in Scopus (green).**

*Timeline Visualization*

Figure 15 shows an overview of the temporal patterns of major clusters formed by references cited in publications in $S_B$. Each cluster is shown as a stream from left to right. The overall color of a cluster represents the time when co-citation connections were made for the first time in the dataset. For example, the cluster on the top contains co-citation arcs in yellow and orange, which correspond to the most recent years in the time frame. The color patterns indicate a few streams in earlier years with references of strong citation bursts (tree rings in red). The earliest one is Cluster #4, followed by a small cluster Cluster 24 located near the bottom, then Cluster #3, and Cluster #1. Clusters emerged in the middle of the time frame include Cluster #5, #9, and #17. In particular, the largest circle of Garfield's 2006 JAMA article appeared in #9, which is labeled as citation analysis. There are four clusters that are currently active, namely, #0 on fractional counting, #2 comprehensive review, #7 PhD student, and #11 science citation index. Cluster labels are chosen from titles of articles that cite members of corresponding clusters. In the following discussion, we will highlight major contributors in the four currently active areas of research.

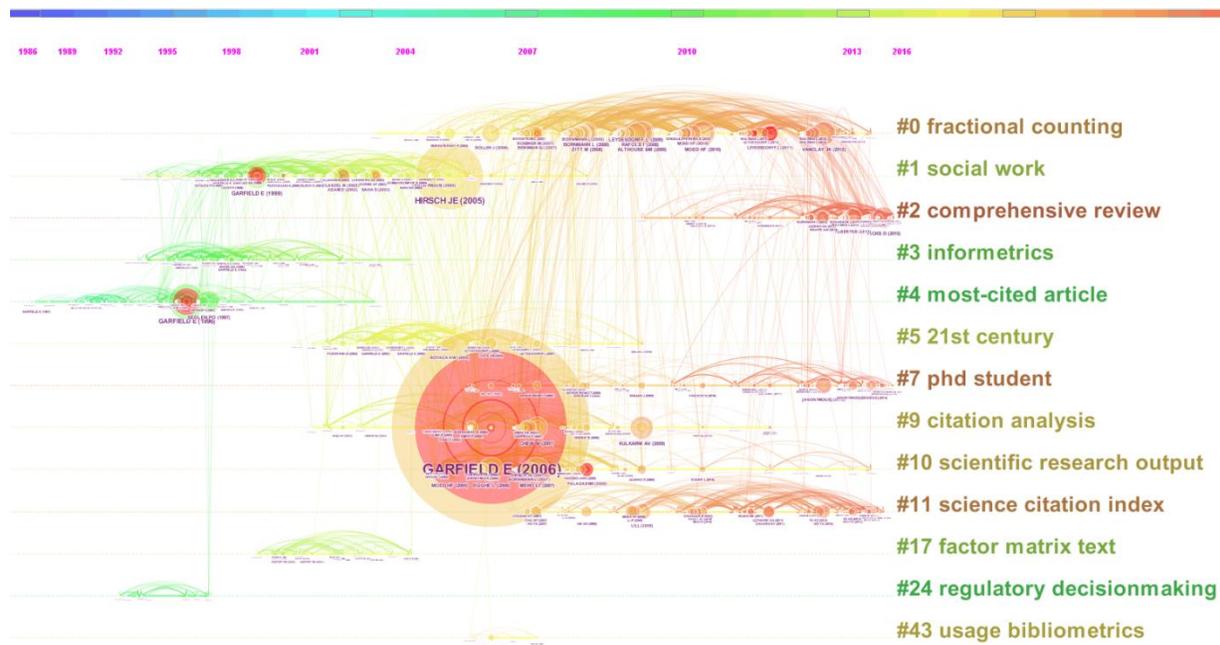

**Figure 15. Timeline visualization of the co-citation clusters.**

*Clusters*

Two of the largest currently active clusters are reviewed as follows, namely, Clusters #0 and #2. In particular, we decompose Cluster #0 further for two more levels below the current level, i.e. Levels 1-2-3.

**Cluster #0: Fractional Counting**

The major theme in citing articles of this cluster is fractional counting and normalization. As shown in the timeline visualization, this cluster spans over 10 years since 2004. The three articles that cited the most members of this cluster are as follows. The first two from Leydesdorff in 2011 are about normalizing citation counts.

> 1. 0.06 Leydesdorff, L (2011) how to evaluate universities in terms of their relative citation impacts: fractional counting of citations and the normalization of differences among disciplines http://dx.doi.org/10.1002/asi.21511
> 2. 0.06 Leydesdorff, L (2011) turning the tables on citation analysis one more time: principles for comparing sets of documents http://dx.doi.org/10.1002/asi.21534
> 3. 0.06 Vanclay, JK (2012) impact factor: outdated artefact or stepping-stone to journal certification? http://dx.doi.org/10.1007/s11192-011-0561-0

Table 13 lists some of the most cited references in Cluster #0. We have seen three of them with citation bursts over 6 years, shown in shaded rows in the table. The cluster at this level of granularity, however, contains a considerable amount of heterogeneity, which means they may represent different semantic groupings even though they share similar citation patterns. We need to decompose the cluster for further inspection.

**Table 13. Most cited references in Cluster #0.**

| Citation | Author | Year | Source | Volume | Page | Half-Life |
|---|---|---|---|---|---|---|
| 50 | Althouse BM | 2009 | J AM SOC INF SCI TEC | 60 | 27 | 3 |
| 48 | Vanclay JK | 2012 | SCIENTOMETRICS | 92 | 211 | 1 |
| 48 | Moed HF | 2010 | J INFORMETR | 4 | 265 | 3 |
| 47 | Rafols I | 2009 | J AM SOC INF SCI TEC | 60 | 1823 | 3 |

| 41 | Zitt M | 2008 | J AM SOC INF SCI TEC | 59 | 1856 | 4 |
| 40 | Leydesdorff L | 2009 | J AM SOC INF SCI TEC | 60 | 348 | 3 |
| 40 | Bornmann L | 2008 | J DOC | 64 | 45 | 4 |
| 36 | Bollen J | 2006 | SCIENTOMETRICS | 69 | 669 | 4 |
| 35 | Bornmann L | 2008 | ETHICS SCI ENV POLIT | 8 | 93 | 4 |
| 34 | Leydesdorff L | 2011 | J AM SOC INF SCI TEC | 62 | 217 | 2 |
| 34 | Moed HF | 2010 | JOURNAL OF INFORMETRICS | 4 | 265-277 | 3 |
| 33 | Falagas ME | 2008 | ARCH IMMUNOL THER EX | 56 | 223 | 4 |
| 32 | Radicchi F | 2008 | P NATL ACAD SCI USA | 105 | 17268 | 4 |
| 32 | Leydesdorff L | 2008 | J AM SOC INF SCI TEC | 59 | 278 | 2 |
| 32 | Falagas ME | 2008 | FASEB J | 22 | 2623 | 3 |
| 31 | Bensman SJ | 2007 | ANNU REV INFORM SCI | 41 | 93 | 4 |

The same visual analytic procedure can be repeatedly applied to each individual cluster. Figure 16 shows a timeline visualization of Cluster #0. We can analyze Cluster #0 at the level of a finer granularity. Now we can see that the decade-long Level 1 Cluster #0 consists of several component streams at Level 2. We may use the notation of #C1#C2 for a Level-1 cluster C1's Level-2 cluster C2. Thus, #0#0 refers to the first sub-cluster of the parent Cluster #0. The parent cluster ID can be omitted if it is clear in context.

The large red arrow in Figure 16 illustrates the relationship between a cluster's labels in a particular year and the cited references in the cluster. The cluster #0#0 is labeled as #0 journal citation report. For each particular year, the most representative label for the cluster would be different based on a subset of publications. Two groups of labels are shown. $LSI_1$ means that labels in that group such as *open access* are from the largest dimension of the Latent Semantic Indexing (LSI) model constructed from the subset. $LSI_2$ means that the corresponding labels are from the second largest dimension, for example, *plagiarism*. This visualization reveals a rich set of information. For example, which sub-cluster includes Garfield's 2006 JAMA paper? Which sub-cluster contains the h-index paper? Which sub-cluster has the same label as the parent cluster?

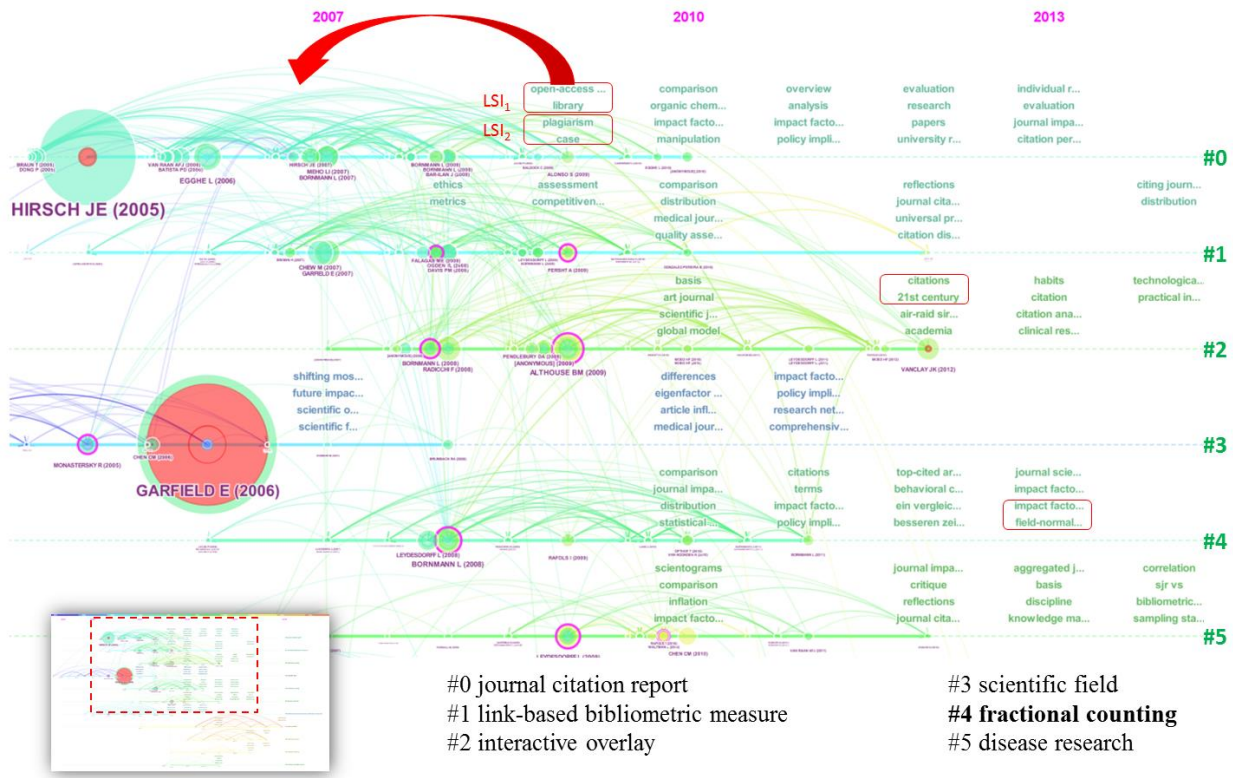

Figure 16. A timeline visualization of Cluster #0, which is further divided into multiple timelines.

The two most recent sub-clusters of Cluster #0 are shown in Figure 17, #0#7 and #0#8. Repeatedly applying the same procedure to a subset of the current data is a technique known as drill-down in information visualization. It is particularly useful for the analyst to explore a cluster with a relatively high degree of heterogeneity, which can be measured in terms of its silhouette score. Cluster #0's silhouette score is 0.742, whereas Cluster #1 has 0.800 and Cluster #2 has 0.909. The heterogeneity of a cluster suggests it may have a complex structure at a deeper level.

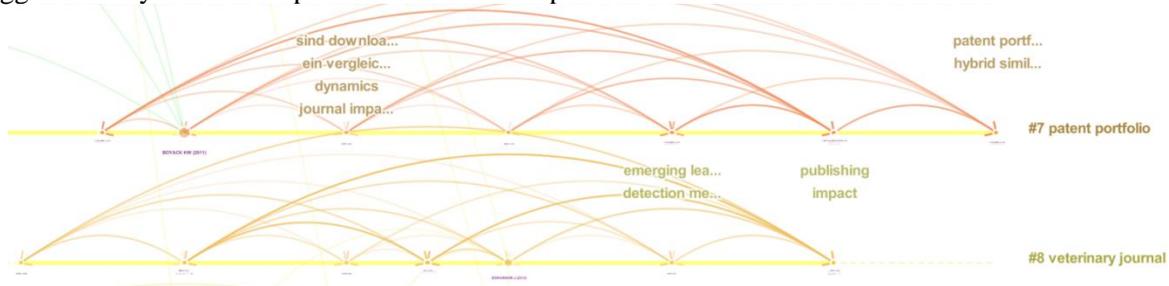

**Figure 17. The most recent sub-clusters of Cluster #0.**

We decomposed the Level 2 cluster #0, i.e. #0#0, one more time and obtained four Level 3 clusters. The silhouette score of each cluster is very high, suggesting a high-level of homogeneity (Table 14). A timeline view of the Level 3 clusters is shown in Figure 18.

**Table 14. Level 3 clusters of the top-level Cluster #0, i.e. Cluster #0#0#0.**

| Cluster ID | Size | Silhouette | Average(Year) | LSI | LLR |
|---|---|---|---|---|---|
| 0 | 40 | 0.987 | 2008 | open-access journals | journal |
| 1 | 21 | 0.959 | 2008 | journal | journal |
| 3 | 15 | 0.894 | 2007 | biohydrogen | scientometric approach |
| 6 | 8 | 0.925 | 2007 | factors | psychology |

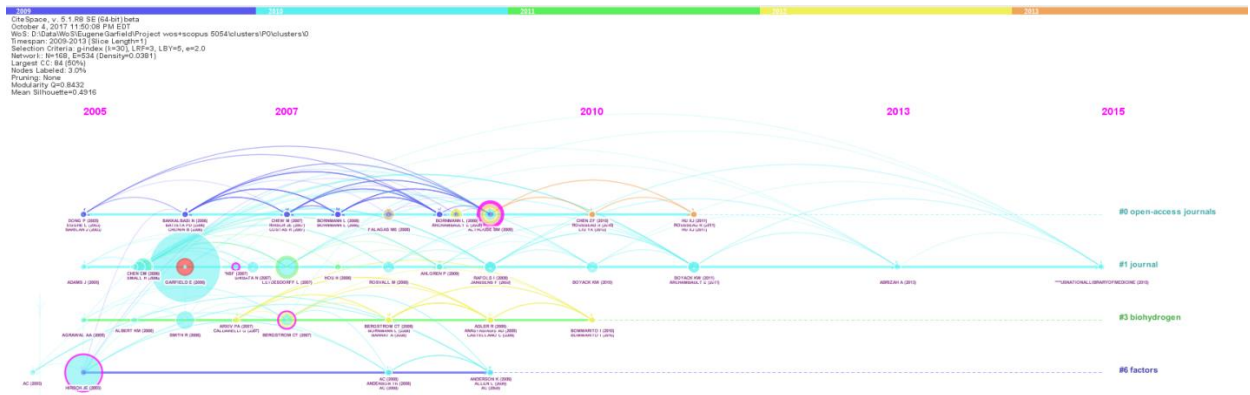

**Figure 18. A timeline visualization of Level 3 clusters of the top-level Cluster #0.**

**Cluster #2: Measuring Research Impact**

The second currently active thread is Cluster #2. The log-likelihood ratio algorithm labels it as comprehensive review. After reviewing its content, its common theme is about measuring research impact, although the algorithm is right in that two of the top three representative articles are literature reviews. The timeline visualization shows that this is a relatively young thread. The three most representative articles are published in 2015 or 2016, including one article on altmetrics.

1. 0.06 Waltman, L (2016) a review of the literature on citation impact indicators http://dx.doi.org/10.1016/j.joi.2016.02.007

2. 0.06 Barnes, C (2015) the use of altmetrics as a tool for measuring research impact http://dx.doi.org/10.1080/00048623.2014.1003174
3. 0.06 Tahamtan, I (2016) factors affecting number of citations: a comprehensive review of the literature http://dx.doi.org/10.1007/s11192-016-1889-2

Most cited references in this cluster are shown in Table 15. This most prominent one is by Hicks et al. (2015) – the Leiden Manifesto for research metrics, which aims to codify ten principles for research evaluation. This is perhaps the most important documentation, aiming at sending clear messages to researchers and practitioners who may be not fully aware of the pitfalls and biases under the seemingly objective disguise of reducing complex phenomena to over simplified numbers.

Table 15. Most cited references in Cluster #2, the second largest currently active Level 1 cluster.

| Citation | Author | Year | Source | Volume | Page | Half-Life |
|---|---|---|---|---|---|---|
| 40 | Hicks D | 2015 | NATURE | 520 | 429 | 1 |
| 35 | Alberts B | 2013 | SCIENCE | 340 | 787 | 2 |
| 27 | Wilhite AW | 2012 | SCIENCE | 335 | 542 | 3 |
| 25 | Lozano GA | 2012 | JOURNAL OF THE AMERICAN SOCIETY FOR INFORMATION SCIENCE AND TECHNOLOGY | 63 | 2140-2145 | 3 |
| 14 | Bohannon J | 2013 | SCIENCE | 342 | 60 | 2 |
| 14 | Waltman L | 2013 | J INFORMETR | 7 | 272 | 2 |
| 13 | Bornmann L | 2012 | RHEUMATOL INT | 32 | 1861 | 3 |
| 13 | Ke Q | 2015 | P NATL ACAD SCI USA | 112 | 7426 | 1 |
| 13 | Wang DS | 2013 | SCIENCE | 342 | 127 | 3 |
| 12 | van Eck N | 2014 | J INFORMETR | 8 | 802 | 2 |
| 12 | Li J | 2012 | SCIENTOMETRICS | 92 | 795 | 4 |

## Discussions and Conclusions

As we integrated datasets from the Web of Science and Scopus, we observed that records from Scopus appear to have a slightly higher level of citation counts. We divided articles into several bins of the same citation counts. In particular, W5 is the set of articles from the Web of Science with citations in the range between 5 and 10 (excluding the highest value in each bin). Figure 19 shows that $Wk \leq Sk$ for k = 5, 10, 20, and 30. Then citation counts from the two sources converge. Further investigations are necessary so that data fusion with multiple sources may consider appropriate normalization procedures as well as field normalization.

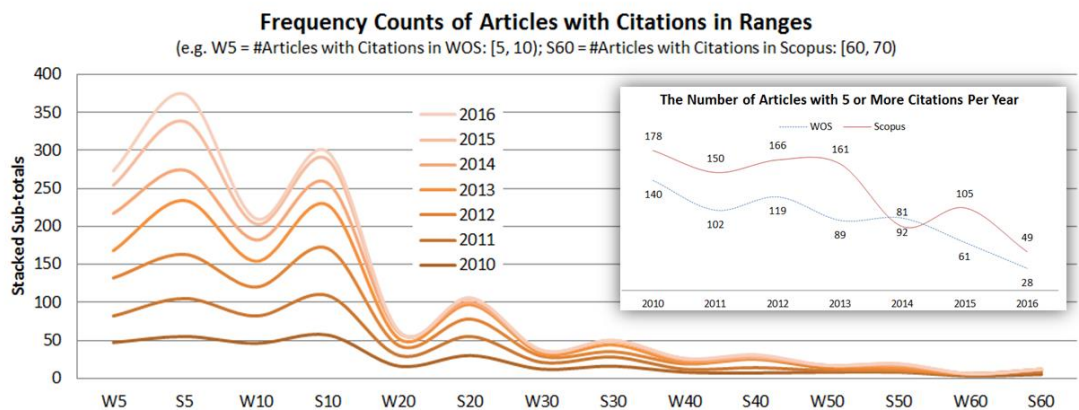

**Figure 19. Records in our sample from Scopus appear to have higher citation counts than their counterparts in the Web of Science.**

In summary, we have visualized and analyzed two datasets $S_A$ and $S_B$. One features Garfield's publications and the other contains publications that cited Garfield's publications. The breadth and depth of Garfield's scholarly impact is demonstrated based on bibliographic records from two major sources, the Web of Science and Scopus. We pay a tribute to Garfield with visual analytic studies of structural and temporal patterns revealed through citation-related connections. Without his invention of citation indexing, none of the studied demonstrated here would be possible. Indeed, much of the research that has built on citations and the rich information that one can learn from citation behaviors may not become the reality. We might save our worries about tools that may fall into the wrong hands, but we may not have valuable tools for the right hands either.

**Acknowledgements**
I'd like to thank Lutz Bornmann for kindly sharing the set of 1,558 records for the study.